\documentstyle[multicol,aps,prl,epsf]{revtex}

\begin{document}

\draft

\title{Fabrication of stable Pd nanowire assisted by hydrogen in solution}

\author{Manabu Kiguchi and Kei Murakoshi}

\address{Department of Chemistry, Graduate School of Science, 
Hokkaido University, Sapporo, 060-0810, Japan}

\date{\today}

\maketitle

\begin{abstract}

We have mechanically fabricated a Pd nanowire in solution under electrochemical potential 
control. A clear feature appeared in the conductance histogram when the electrochemical 
potential of the Pd wire was kept at the hydrogen evolution potential. Conductance traces 
showed the Pd wire was stretched 0.4 nm in length just before breaking, suggesting that at 
least two Pd atoms might contribute to the formation of the Pd wire. The results indicate that a 
certain atomic configuration of the Pd nanowire is stabilized by hydrogen. We discuss the 
stabilization mechanism due to changes in bond strengths caused by hydrogen adsorption or 
incorporation.

\end{abstract}

\medskip

\begin{multicols}{2}
\narrowtext

Fabrication and characterization of metal atomic wires have attracted attention due to their 
potential applications in future electronic devices \cite{1}. Recently, Ir, Pt, Au mono atomic wires 
have been fabricated by pulling nano contacts using a scanning tunneling microscope (STM) 
or a mechanically controllable break junction (MCBJ) in ultra high vacuum (UHV) at low 
temperatures \cite{1,2,3}. In addition to methods in UHV, the electrochemical method has been 
recognized as another powerful approach in the fabrication of metal nanowires \cite{4,5,6,7,8}. Recently, 
we have demonstrated that Fe, Co, and Ni metal nanowires were stabilized in solution at room 
temperature via the hydrogen evolution reaction \cite{6,7}. Furthermore, we observed results 
suggesting the formation of a mono atomic Ni wire in solution, while mono atomic wires of 
3d and 4d metals have not been prepared in UHV up to now.

In the present study, we have studied the hydrogen assisted stabilization of the Pd (4d 
metal) nanowire in the fabrication of a mono atomic wire. A mono atomic Pd wire is predicted 
to be ferromagnetic \cite{9}, and to test this hypothesis it is useful to fabricate this wire. In addition 
to the stabilization of the Pd nanowire by hydrogen, the interaction between hydrogen and the 
Pd nanowire should be interesting. Since the interaction between hydrogen and Pd is large, 
the nanowire of the palladium hydride formed in solution should have electrical properties \cite{10}.

In our study of the Pd nanowire, we have chosen to measure the conductance. Electrical 
conductance through a metal nanowire on an atomic scale is expressed by $G=2e^{2}/h \sum T_{i}$ 
 where $T_{i}$  is the transmission probability of the $i$-th conductance channel, $e$  is the 
electron charge, and $h$  is Plank's constant \cite{1}. Since conductance quantization depends on the 
atomic structure of a nanowire and the inherent properties of metals, we can study the atomic 
and electronic structures of metal nanowires by the conductance measurements. We have 
already studied conductance quantization of a Pd nanowire in solution \cite{8}. The conductance 
histogram displayed a clear feature peak at 1 $G_{0}$, while conductance quantization was not 
observed in UHV at room temperature \cite{11}. At low temperatures in UHV, the appearance of the 
1 $G_{0}$ and 0.5 $G_{0}$ features were reported in the conductance histogram for Pd under hydrogen 
dosing \cite{12}. At the present stage, it is clear that hydrogen stabilizes the structure with a 1 $G_{0}$ or 
0.5 $G_{0}$ both in UHV and solution. However, there has been little discussion about the structure, 
that is, it is not clear whether the conductance originates from the bridging hydrogen 
molecules or the Pd nanostructure stabilized by hydrogen.

In the present study, we have investigated the mechanical properties of the Pd nanowire 
under electrochemical potential control. To obtain information about the structure of the 
nanowire, we have analyzed each conductance trace, as the conductance traces provide 
information about the nanowire such as wire length and thickness. Using the trace analysis 
results, we discuss the formation mechanism of the Pd mono atomic wire in solution. 

The experimental design used in this study was the same as described in our previous 
report \cite{5,6,7,8}. An electrochemical scanning tunneling microscope (EC-STM) was used to fabricate 
the Pd nanowire. The electrochemical potential ($\Phi_{0}$) was controlled using a potentiostat 
(Pico-Stat, Molecular Imaging Co.) with a Ag/AgCl reference electrode. The electrolyte 
consisted of 1 mM PdCl$_{4}$, 2 mM HCl and 0.05 M H$_{2}$SO$_{4}$. The Pd nanowire was prepared as 
follows. First, the electrochemical potential of both the STM tip and substrate were 
maintained at less than +450 mV where Pd bulk deposition proceeds \cite{13}. After the sufficient 
deposition of Pd onto the surfaces of the STM tip and the substrate, the tip was pressed into 
the substrate and contacts were then pulled out. During contact breaking, a Pd nanowire is 
formed between the tip and substrate. Conductance was measured during the breaking process 
under an applied bias of 20 mV between the tip and substrate. Statistical data (conductance 
histograms, etc.) was obtained for a large number (over 3000) of individual conductance 
traces.

Figure 1 (a) shows the cyclic voltammogram of the Au electrode in solution. Pd deposition 
and hydrogen adsorption/evolution proceeded when the electrochemical potential of Au 
electrodes were maintained under more negative conditions than $\Phi_{0}$= 450 mV and $\Phi_{0}$= -200 
mV, respectively \cite{13,14,15}. Figure 1 (b) and (c) show the conductance histogram of the Pd 
nanowire at (b) $\Phi_{0}$= -250 mV and (c) $\Phi_{0}$= 100 mV. There is no clear feature in the 
conductance histogram at $\Phi_{0}$= 100 mV, suggesting that no preferential atomic configurations 
were formed. The featureless conductance histogram agreed with the previous results in UHV 
at room temperature \cite{11}. When the electrochemical potential of the nanowire was maintained 
under more negative conditions ($\Phi_{0}$= -200 mV), a feature appeared near the unit of the 
quantum conductance (0.9 $G_{0}$). These results agree with our previous results \cite{8}.

Next, we analyzed the conductance traces to determine the structure of the Pd nanowire. 
Figure 2 shows the typical conductance traces at (a) $\Phi_{0}$= 100 mV and (b,c) $\Phi_{0}$= -250 mV. As 
a general tendency, the conductance changed in a stepwise fashion rather than continuously. 
At $\Phi_{0}$= 100 mV, the conductance values of the plateaus were non-reproducible, resulting in 
featureless conductance histograms (Fig. 1 (c)). On the other hand, the 0.6-1.2 $G_{0}$ plateau with 
a stretched length ca. 0.5 nm appeared in the trace at $\Phi_{0}$= -250 mV as shown in Fig. 2(b). 
These characteristic plateaus were the origin of the feature at 0.9 $G_{0}$ in the conductance 
histogram (Fig. 1 (b)). The value of 0.9 $G_{0}$ may correspond to that of a Pd mono atomic 
contact. This value is comparable to previously documented estimations of clean transition 
metal mono atomic contacts in UHV systems \cite{1}. The length of the plateau also showed 
dependence on the electrochemical potential. To evaluate characteristics of the plateau length 
quantitatively, we examined the distribution of lengths for the last conductance plateau. For 
comparison, we examined the Pt nanowire under the hydrogen evolution, in which a hydrogen 
molecule would bridge between Pt electrodes. Smit et al. showed that the 1 $G_{0}$ feature 
appeared in the conductance histogram under hydrogen dosing at low temperatures in UHV \cite{16}. 
By measuring the conductance and vibration spectra, they showed that the 1 $G_{0}$ feature 
originated from a single hydrogen molecule bridging Pt electrodes. In the breaking trace, the 1 
$G_{0}$ plateau broke within 0.2 nm, which is consistent with the small size of the hydrogen 
molecule \cite{16}. The inset of Fig. 3 (a) shows the conductance histogram of the Pt nanowire under 
hydrogen evolution reaction in solution. A feature close to 1 $G_{0}$ appears in the histogram. The 
feature appears to have originated from the bridging hydrogen molecule, although the 
situation in solution might be different from that in UHV at low temperatures. Figure 3 (a) 
shows the distribution of lengths for the last conductance plateau for Pt. The length of the last 
plateau was defined as the distance between the points at which the conductance dropped 
below 1.3 $G_{0}$ and 0.7 $G_{0}$, respectively. These values were fixed in the following analysis. The 
contact broke at a short length of ~0.2 nm which is in agreement with previous results \cite{16}.

Figure 3 (b) shows the distribution of lengths for the last conductance plateau for Pd at $\Phi_{0}$= 
-250 mV. For comparison, we also show the distribution of the lengths for Pd at $\Phi_{0}$= 100 mV. 
While the contact broke within 0.2 nm at $\Phi_{0}$= 100 mV, the plateau was stretched 0.4 nm in 
length at $\Phi_{0}$= -250 mV. If a hydrogen molecule bridges between the Pd nanowire, the plateau 
length would be restricted to be 0.2 nm using the Pt results. The long plateau suggests that the 
1 $G_{0}$ feature did not originate from the hydrogen bridges. Since the Pd-Pd distance is 0.27 nm 
for bulk Pd, the 0.4 nm long 1 $G_{0}$ plateau corresponds to a mono atomic wire of about two Pd 
atoms. A hydrogen atom would adsorb on the surface or be incorporated in the wire under the 
hydrogen evolution reaction, and thus, a stable Pd nanowire would be formed in solution.

The formation of a Pd mono atomic wire was also supported by the conductance trace. We 
occasionally observed reversible transition of conductance between a 1 $G_{0}$ and 0.6-0.7 $G_{0}$ as 
seen Fig. 2(c). A similar reversible transition of conductance was reported for Au mono 
atomic wires in the presence of physically adsorbed hydrogen in UHV at low temperatures 
(10-30 K) \cite{17,18}. Similar behavior was also observed for the Au mono atomic wire in solution 
under electrochemical potential control \cite{19}. The fractional conductance and reversible 
transition are explained by the formation of a dimerized structure in the Au mono atomic wire, 
and the dynamic structural transition between a dimerized wire and an equal-spacing wire, 
respectively. The dimerization is a characteristic of a mono atomic wire. Therefore, the 
reversible transition of the conductance observed in the present study may reflect a similar 
dynamic structural transition in the Pd mono atomic wire. Present observations strongly 
suggest that the Pd mono atomic wire was fabricated in solution at room temperature under 
the hydrogen evolution reaction. Hydrogen stabilized the specific structure of the Pd mono 
atomic wire in solution at room temperature, leading to the characteristic conductance 
quantization behavior which has not observed in UHV at room temperature. At the present 
stage, it remains unclear whether hydrogen atom adsorbs on the wire or incorporates in the 
wire. In-situ point contact spectroscopy should be indispensable to obtain an information on 
the structure \cite{16}.

Adsorbed or incorporated hydrogen may play a decisive role in improving the stability of 
Pd mono atomic wire in solution. Generally, the bond strength between atoms increases when 
the coordination number of neighboring atoms becomes low \cite{3}. Further increments in the 
stabilization becomes significant in 5d metals due to the relative effects of the valence 
electron, leading to the formation of a very stable nanowire of 5d metals. In the case of 3d and 
4d metals, it is known that stabilization of the bond strength due to changes in the 
coordination number as well as from the relative effect of electrons is not significant. These 
characteristics are key in the difficulty of preparing 3d and 4d metal nanowires showing stable 
conductance quantization. However, hydrogen adsorption/incorporation in the Pd nanowire is 
expected to change this situation. Adsorption of hydrogen to Pd proceeds via electron transfer 
from Pd to hydrogen \cite{20}. Since the top of the 4d band of bulk Pd consists of states with 
anti-bonding characteristics, a decrease in the occupancy of the anti-bonding 4d band due to 
hydrogen adsorption would increase the bond strength between Pd atoms. Thus, hydrogen 
adsorption/incorporation may result in the stabilization of the Pd nanowire. This change in the 
electronic structure of the Pd nanowire under the hydrogen evolution reaction resulted in the 
fabrication of the stable Pd nanowire with 0.9 $G_{0}$. Significant changes in the conductance 
value from 1.8 $G_{0}$ clean Pd to 0.9 $G_{0}$ hydrogen adsorption/incorporation Pd may also originate 
from this change in the electronic structure of the Pd nanowire. Finally, we briefly compare 
the present results with that in UHV at low temperature. While the 1 $G_{0}$ and 0.5 $G_{0}$ features 
were reported in the conductance histogram for Pd in UHV at low temperature \cite{12}, the 0.5 $G_{0}$ 
feature was not observed in the conductance histogram under hydrogen evolution reaction in 
solution. In UHV, a hydrogen molecule is introduced to the Pd nano wire. On the other hand, 
in solution under hydrogen evolution reaction, an atomic hydrogen is formed on the Pd 
surface as dominant intermediate of the reaction \cite{21}. The situation in solution, which is 
different from that in UHV at low temperature, may lead to the absence of the structure of 0.5 
$G_{0}$ peak in the histogram.

In conclusion, we have observed conductance quantization behavior of mechanically 
fabricated Pd nanowire under electrochemical potential control. Under the hydrogen evolution 
reaction, a peak close to a 1 $G_{0}$ appeared in the conductance histogram, suggesting that a 
certain atomic arrangement was maintained in solution. Relatively long 0.4 nm stretch lengths 
and reversible conductance fluctuations during stretching suggests the formation of a 
one-dimensional Pd mono atomic wire in solution at room temperature. 
This work was partially supported by a Grant-in-Aid for Scientific Research A (No. 
16205026) and Grant-in-Aid for Scientific Research on Priority Areas (No. 17069001) from 
MEXT.

\begin{figure}
\begin{center}
\leavevmode\epsfysize=55mm \epsfbox{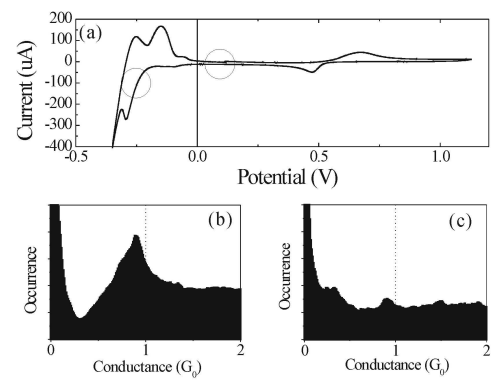}
\caption{(a) Cyclic voltammogram observed in 1 mM PdCl$_{4}$, 2 mM HCl and 0.05 M H$_{2}$SO$_{4}$. 
The Conductance histograms at (b) $\Phi_{0}$ = -250 mV, (c) $\Phi_{0}$ = 100 mV.   }
\label{fig1}
\end{center}
\end{figure}

\begin{figure}
\begin{center}
\leavevmode\epsfysize=50mm \epsfbox{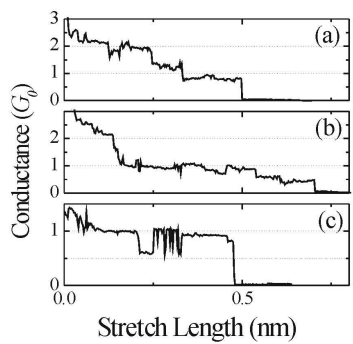}
\caption{Typical conductance traces of Pd nanowire at (a) $\Phi_{0}$ = 100 mV, (b,c) $\Phi_{0}$ = - 250 mV.}
\label{fig2}
\end{center}
\end{figure}

\begin{figure}
\begin{center}
\leavevmode\epsfysize=50mm \epsfbox{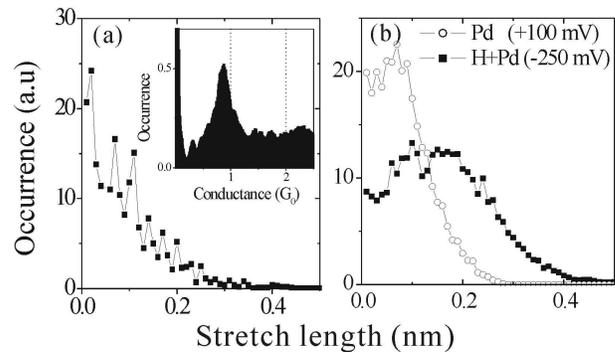}
\caption{The distribution of lengths for the last conductance plateau for Pt under hydrogen 
evolution reaction ($\Phi_{0}$ = - 800 mV). Inset: Conductance histogram of Pt nanowire at $\Phi_{0}$ = - 800 
mV. (b) Distribution of lengths for the last conductance plateau for Pd at $\Phi_{0}$ = 100 mV and $\Phi_{0}$ = 
- 250 mV. }
\label{fig3}
\end{center}
\end{figure}

\end{multicols}
\end{document}